\documentclass{iopart}
\usepackage{epsfig}
\usepackage{hyperref}
\begin{document}

\title{Experimental Single-Impulse Magnetic Focusing of Launched Cold Atoms}
\author{David A Smith\dag,
\href{http://www.photonics.phys.strath.ac.uk/People/Aidan/Aidan.html}{Aidan S Arnold}\ddag, Matthew J Pritchard\dag~and
\href{http://massey.dur.ac.uk/igh/index.html}{Ifan G Hughes}\dag}
\address{\dag~Department of Physics, Rochester Building, University of
Durham, South Road, Durham, DH1 3LE, UK}
\address{\ddag~SUPA, Department of Physics, University of Strathclyde,
Glasgow, G4 0NG, UK} \ead{i.g.hughes@durham.ac.uk}
\date{\today}
\begin{abstract}
Single-impulse three-dimensional magnetic focusing of vertically launched cold atoms has been observed.  Four different configurations of the lens were
used to vary the relative radial and axial focusing properties. Compact focused clouds of $^{85}$Rb were seen for all four configurations.  It is shown
that an atom-optical ray matrix approach for describing the lensing action is insufficient. Numerical simulation using a full approximation to the lens's
magnetic field shows very good agreement with the radial focusing properties of the lens. However, the axial (vertical direction) focusing properties are
less well described and the reasons for this are discussed.

\end{abstract}
\pacs{03.75.Be, 32.80.Pj, 39.25.+k}

\section{Introduction}
Laser-cooled atoms \cite{nobel97} are  extensively used in a range of experiments spanning fundamental physics and technological applications. As the
kinetic energy of ultracold atoms is many orders of magnitude lower than conventional atomic beams, relatively modest electromagnetic forces are now
routinely used to gain complete control over the external degrees of freedom of atomic motion \cite{adams97, Hinds}; these developments heralded an
upsurge of interest in the field of atom optics~\cite{Adams94}. One of the goals in this field  is to realise atom-optical elements that are analogues of
conventional optical devices, such as mirrors and  lenses. An atom mirror reverses the component of velocity perpendicular to the surface and maintains
the component parallel to the surface, whereas an atom lens can modify both the transverse and  longitudinal velocity components. In addition to
atom-light interactions the Stern-Gerlach force has been used to realise flat atomic mirrors~\cite{flat}, curved atomic mirrors~\cite{curved}, and pulsed
mirrors for both cold (thermal)~\cite{kadio} and Bose-condensed atoms~\cite{BECmirror}.

There are numerous reasons for studying the focusing of cold atoms using lenses or curved mirrors, including: transferring cold atoms from a
magneto-optical trap (MOT) to a spatially-separated vacuum chamber of lower background pressure~\cite{Szymaniek99}; atom lithography~\cite{meschede}; or
loading miniature magnetic guides~\cite{minmagguide} and atom chips~\cite{atomchips}. In comparison to an unfocused cloud, the atom density  can be
significantly increased after magnetic focusing.

Pulsed magnetic lenses for 3D atom focusing were first demonstrated by Cornell~{\it et al.}~\cite{mon1} using the alternate-gradient technique. The group
of Gorceix has performed experiments demonstrating the longitudinal Stern-Gerlach effect with an atomic cloud using pulsed magnetic forces \cite{Marec},
and an experimental and theoretical study of cold atom imaging by alternate-gradient magnetic forces \cite{Gor}. Previously we have studied the
theoretical performance of both single \cite{focus1} and double-impulse focusing \cite{focus2}. In this paper we demonstrate  a single-impulse strategy
with a baseball lens.  The advantages of this scheme are the simplicity of the design, and the theoretical prediction that this scheme is ideal for
achieving the goal of minimizing the root-mean-square (rms) image volume of a launched cloud~\cite{focus2}.

The remainder of the paper is organised as follows: Section~\ref{baseball}  outlines the theory and construction  of the baseball lens;
Section~\ref{exptl}  discusses the experimental details; in Section~\ref{results} the results are presented and analysed; finally, in
Section~\ref{conclusions}, conclusions are drawn.  The reader interested in the theory of atom focusing with pulsed magnetic fields is referred to
earlier work \cite{Gor, focus1, focus2}.

\section{The Baseball Lens -- construction and characterization}\label{baseball}
As it is impossible to create a static magnetic field maximum \cite{earnshaw}, there is only one strategy for producing a focused cloud with a single
magnetic impulse -- one uses atoms in weak-field-seeking states, and a lens potential with a minimum at the centre and positive curvature along all three
Cartesian axes. This is essentially the requirement for a magnetic trap, for which many designs exist. A magnetic trap/lens also requires a non-zero
minimum field, to avoid spin-flip losses~\cite{petr}. In~\cite{focus1} we analysed the aberrations expected from different magnetic lenses and concluded
that a baseball lens would be ideal for achieving single-impulse three-dimensional focusing.

The baseball lens is a variant of the Ioffe-Pritchard trap~\cite{IPrefs}. Figure~1 part (a)  shows the geometry.  Previous work with quadrupole and
Ioffe-Pritchard devices (for cold atoms and BECs \cite{kadio,BECmirror}) led to strong focusing in  two dimensions. A novel feature of our work is that
the lens provides an isotropic potential.  The baseball lens used here has two components: a nine-turn baseball coil carrying a current $I',$ and a pair
of two-turn circular bias coils which carry the same current $I$ in the same sense. The baseball coil consists of eight straight current-carrying
segments of length $w=10\,$cm along $x,y$, and $\ell=10\,$cm along $z$. The bias coils have radius $a=5\,$cm and are separated by $s=5\,$cm. The coils
were constructed from enamel-coated $3\,$mm-diameter cylindrical copper. The ratio of the axial and radial magnetic curvatures can be tuned via the
current ratio $I/I'$. The bias field is needed because it is impossible to realise a 3D isotropic lens with a baseball coil alone. Four configurations of
the baseball lens were used in this experiment, i.e. four ratios between the axial and radial curvatures.   The baseball lens was designed to run with
hundreds of Amps for tens of milliseconds; consequently, the coils were not water-cooled.

Three 12~V truck batteries in series provided the current pulse. An integrated gate bipolar transistor (IGBT) was used to control the current pulse, and
a reverse-biased Schottky diode in parallel with the load prevented oscillatory currents in the lens after switch-off. The baseball and bias coils have
resistances of $18\,$m$\Omega$ and $3\,$m$\Omega,$ with impedances of $32\,\mu$H and $2\,\mu$H, respectively.  The turn on(off) time for the current is
$\sim2\,$ms. Further details of the lens construction and circuit can be obtained in~\cite{davethesis}.

The second-order expansion of the magnetic field magnitude of a baseball lens requires five parameters: the axial bias field and field curvature from the
bias coils, and the axial bias field, gradient and curvature  from the baseball coil.  Theoretical expressions for these quantities in terms of the
currents and dimensions of the baseball and bias coils can be found in~\cite{focus1}. These parameters were measured with a 10~A test current, and the
measured and theoretical values are in good agreement~\cite{davethesis}.

\section{Experimental Setup and Procedure}\label{exptl}

\begin{figure}[!b]
\begin{center}
\vspace{-3mm}\epsfxsize=.63 \columnwidth \epsfbox{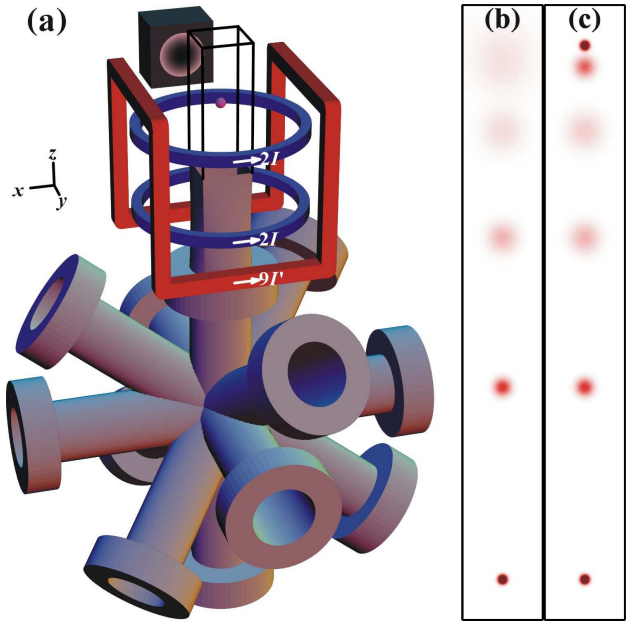} \caption{\label{trajectories+schematic} Image (a) is a schematic of the experimental
apparatus. The MOT is realised at the centre of the intersecting six-way crosses.  After a vertical moving-molasses launch the baseball lens is pulsed on
at the appropriate time to focus the atoms in the quartz cell. The CCD camera (black box) was positioned to capture fluorescent images in the $xz$ plane.
Image (b) shows pictures of a launched cloud at equal time intervals; in the absence of a focusing pulse the cloud volume grows cubically with time, with
a corresponding decrease in atom density. Image (c) shows the effect of a magnetic impulse - the final cloud volume is decreased with a concomitant
increase of density. Parts (b) and (c) are scaled diagrams for the dimensions used in the experiment discussed in this paper. The images saturate (dark
red) at 25\% of the maximum initial density.}\end{center}
\end{figure}

Figure~\ref{trajectories+schematic} (b) and (c) show the principle of the experiment, comparing unfocused and focused atomic trajectories; part (a) is a
schematic of the apparatus. The experiment utilised a custom-made stainless steel vacuum chamber having 12 ports, composed of two intersecting 6-way
crosses.  One cross had two sets of orthogonal ports in the horizontal plane, and one vertical pair.  The other cross had 3 mutually orthogonal axes,
symmetrically disposed about the vertical (at an angle $\cos^{-1}(1/\sqrt{3})$), along which the MOT beams propagated. The advantage of this setup is
that only two laser frequencies are required to achieve vertical moving molasses, and the propagation direction of the atoms is free for a probe beam.
The chamber was pumped with a magnetically-shielded $40\,$l/s ion pump and the background pressure was 9$\times 10^{-11}$~Torr. The centre of a
square-cross-section glass cell was located  $20.5\,$cm above the MOT to enable the focused atoms to be observed. Three pairs of mutually orthogonal
magnetic field coils were used to cancel ambient  fields in the chamber.

A  MOT containing 7$\times 10^7$ $^{85}$Rb atoms was achieved using 6 independent circularly-polarised beams, each of $10\,$mm ($1/e^2$) radius and power
$P=3\,$mW, red-detuned $11\,$MHz from the $^{85}$Rb $5S_{1/2}$ $F=3 \rightarrow 5P_{3/2}$ $F^{'}=4$ transition. Approximately $5\,$mW of repumping light
was shared amongst the 6 MOT beams. The trapping and repumping beams were produced by two
 grating-stabilized, external-cavity diode lasers locked using polarization spectroscopy
\cite{PolSpec} with hyperfine pumping/saturated absorption spectroscopy as a frequency reference \cite{AJP}. Rb vapour was provided by an SAES
dispenser. The magnetic quadrupole field had an axial gradient of $15\,$G/cm.

After collection in the MOT, the atoms underwent a $10\,$ms $28\,$MHz-red-detuned optical molasses stage with 25\% trap laser intensity, which gave a
temperature of (25$\pm$2)~$\mu$K. A frequency difference of $\delta\nu=1.48\,$MHz between the upwards and downwards propagating laser beams then launched
the atoms vertically in moving molasses at a speed of $2.0\,$m/s. The frequency ramp of $\delta\nu$ took $3\,$ms and the final value was held for a
further $1\,$ms. These values were optimised by studying images of the launched cloud up to $20\,$ms after the launch process. The initial cloud standard
deviations were measured to be  $\sigma_x =$1.01$\pm$0.01~mm and $\sigma_z =$0.97$\pm$0.01~mm. After launch, the atoms were optically pumped into the
weak-field-seeking 5$S_{1/2}$ $|F\!=\!3,M_{F}\!=\!3\rangle$ state using a $300\,$mG vertical magnetic field and a $50\,\mu$s pulse of $350\,\mu$W
retro-reflected, vertically-propagating, circularly-polarised light resonant with the $^{85}$Rb $5S_{1/2}$ $F=3 \rightarrow 5P_{3/2}$ $F^{'}=4$
transition. Repumping light resonant with the $^{85}$Rb $5S_{1/2}$ $F=2 \rightarrow 5P_{3/2}$ $F^{'}=3$ transition was present to prevent atoms from
accumulating in undesired states.

Fluorescence images of the launched clouds were taken at the apex of flight ($204\,$ms after launch) using the same beams that were used for optical
pumping, but with  a duration of $2\,$ms  and a  power of $6\,$mW. There is no evidence of significant atom loss during the focussing. We were careful to
ensure that the imaging pulse did not blur or displace the image of the cloud by virtue of the radiation pressure exerted on the atoms. The centre of the
baseball lens was located 16.5$\pm$0.2~cm above the MOT. The unfocused cloud came to rest in (approximately) the centre of the image. For each pulsed
magnetic lens duration, $\tau$, the lens turn-on time was adjusted to centre the focused cloud in the image. The area seen in the image was
($x$=18.1~mm)$\times$($z$=25.8~mm).

\section{Results and Analysis}\label{results}

Four different lens configurations were realised, each with a different ratio between the axial and radial frequencies. The parameters of these lens
configurations (labelled `1,' `2,' `3,' `4') are shown in Table~\ref{tab:currents}.

\begin{table}[!ht]
\begin{center}
\begin{tabular}{|c|c|c|c|c|}
\hline Lens Config. & $I'\,$(A) & $I\,$(A) & $\omega_{x}$ (rad$\,$s$^{-1}$) &
$\omega_{z}$ (rad$\,$s$^{-1}$) \\
\hline \hline
1 & $832\pm4$ & $832\pm4$ & $30\pm1$ & $38\pm2$ \\
2 & $872\pm4$ & $446\pm3$ & $38\pm1$ & $39\pm1$ \\
3 & $898\pm4$ & $304\pm2$ & $41\pm2$ & $39\pm2$  \\
4 & $947\pm5$ & 0 & $50\pm2$ & $40\pm2$ \\\hline
\end{tabular}
\caption{Parameters for different lens configurations. The angular frequencies are deduced from field measurements.  The slight variation of $\omega_{z}$
is due to the varying value of the baseball current $I'$, since the same current is used to feed both the bias coils and the baseball coil. The currents
$I$ and $I'$ were measured with Hall effect current sensors.}
        \label{tab:currents}\vspace{-3mm}
\end{center}
\end{table}

\begin{figure}[!b]
\begin{center}
\epsfxsize=1 \columnwidth\epsfbox{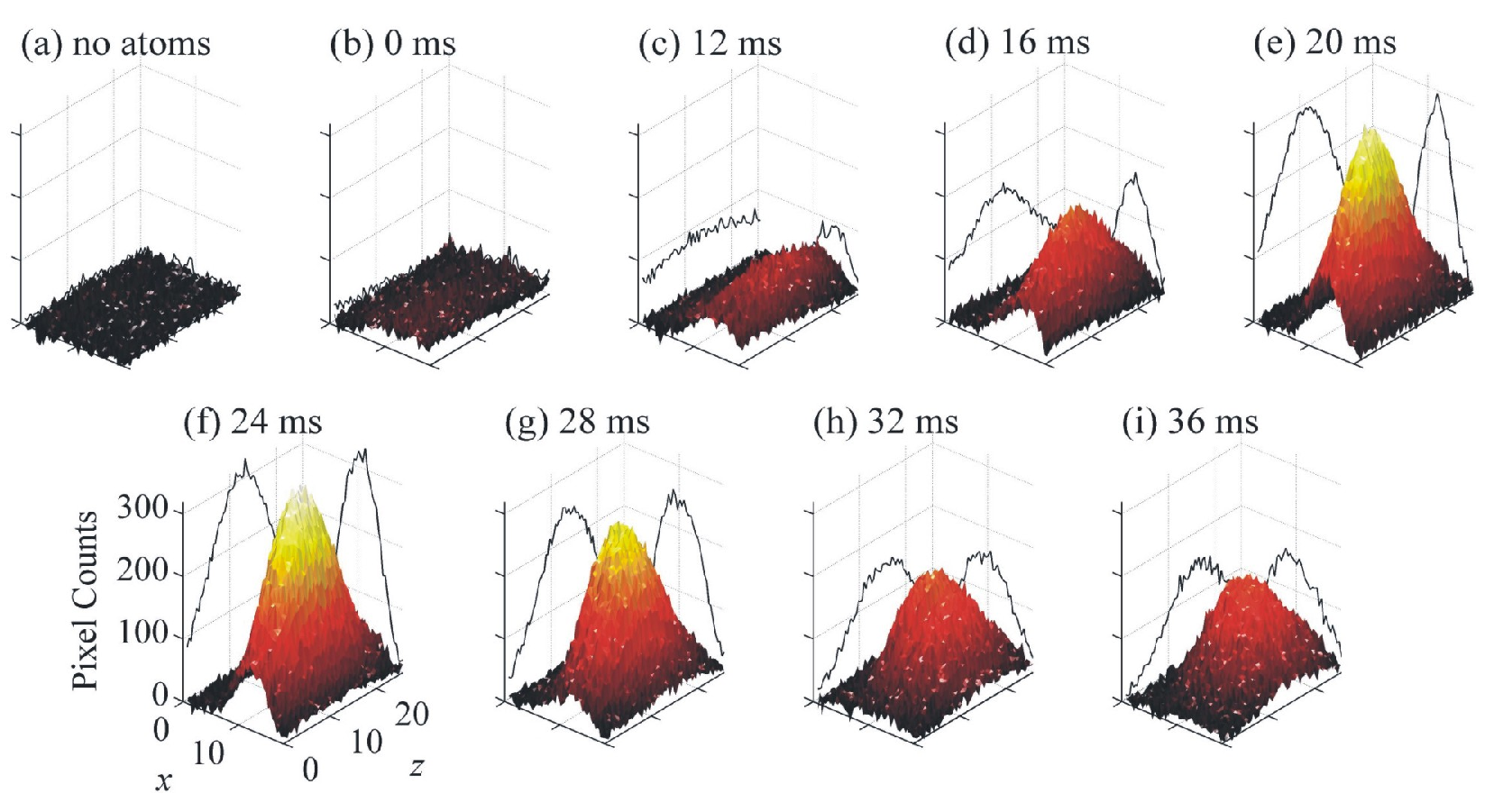} \caption{\label{3Dresults} A sequence of images for increasing baseball lens duration, $\tau$, using lens
configuration 1. (a) Image taken, but no atoms launched; (b) typical image of launched atoms, without lensing ($\tau=0$); (c) - (i) $\tau$ increases from
$12\,$ms to $36\,$ms in 4$\,$ms steps. The $x$ and $z$ axes are in mm.}\end{center}
\end{figure}

Figure~\ref{3Dresults} shows a sequence of images obtained with increasing baseball pulse duration, $\tau$, using lens configuration 1. A background
image with no atoms launched is shown in (a). In (b) a cloud of atoms was launched but not focused.  For the launch temperature, it is expected that the
width of the unfocused Gaussian cloud is significantly larger than the area imaged onto the CCD chip. Images (c) - (i) show the variation of the focused
cloud as a function of $\tau$. The cloud comes to a focus in the $x$-direction between 16 and $20\,$ms, and in the $z$-direction between  28 and
$32\,$ms. Three-dimensional focusing of a launched cloud with a single impulse from a baseball lens is clearly seen.

\begin{figure}[!t]
\begin{center}
\epsfxsize=.9 \columnwidth \epsfbox{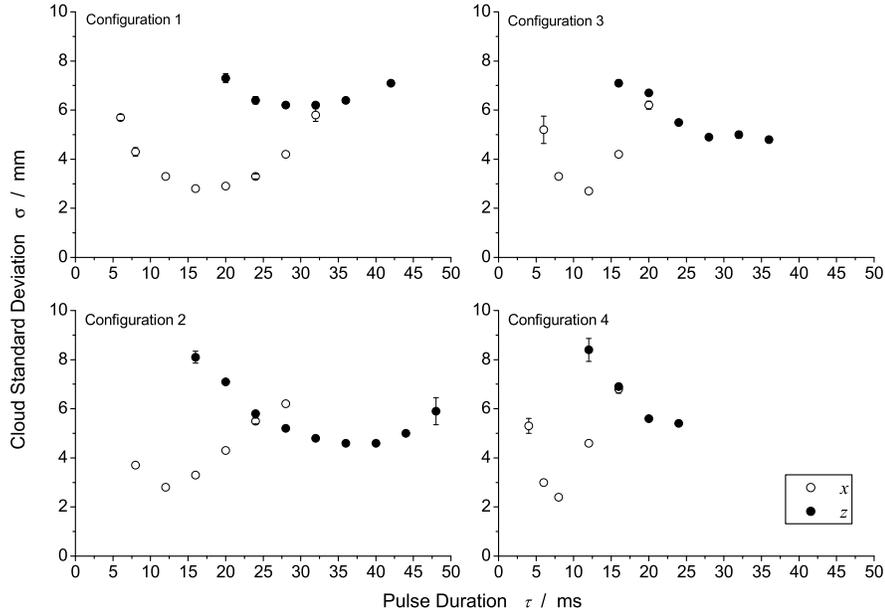}\caption{\label{1stpanel} Plots of the cloud sizes along the $x-$ and $z-$directions, as functions of
pulse duration, $\tau$, for the four lens configurations.  Where the error-bars are not shown they are smaller than the symbols.}\end{center}
\end{figure}

Figure~\ref{1stpanel} shows the cloud sizes (standard deviations) along the $x-$ and $z$-directions for different durations of the impulse, $\tau$, for
all four lens configurations. Three-dimensional magnetic focusing with a single magnetic impulse has been observed for all four configurations, most
notably using lens configuration 1. The radial frequency of a Ioffe-Pritchard trap increases with decreasing bias field.  This is reflected in the data:
as we change from lens configuration 1 to lens configuration 4, the radial angular frequency increases and the pulse duration required to achieve the
minimum $x$-focus decreases. The minimum measured value of the standard deviation in the $x$-direction of a focused cloud was $2.43\pm0.07$\,mm, using
lens configuration 4 with a pulse duration $\tau=8\,$ms. The minimum measured value of the standard deviation in the $z$-direction of a focused cloud was
4.57$\pm$0.03~mm, using lens configuration 3 and a pulse duration of $\tau=36\,$ms. For all four lens configurations, the minimum radial cloud width is
smaller than the minimum axial width.

\subsection{Analysis of cloud size for different lens configurations}
\begin{figure}[!t]
\begin{center}
\epsfxsize=.9 \columnwidth \epsfbox{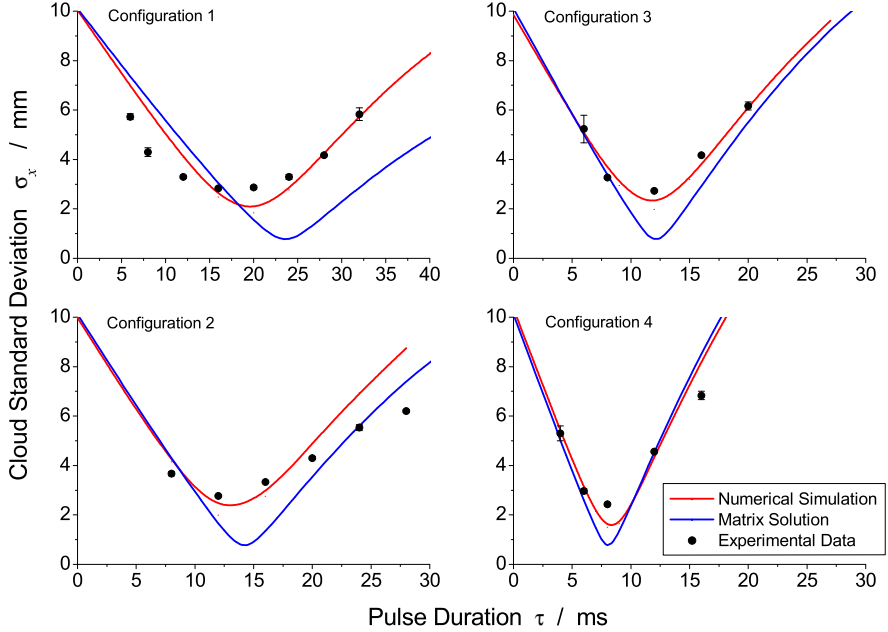} \caption{\label{2ndpanel} Plots of the cloud size along the $x$-direction as a function of the pulse
duration, $\tau$, and comparison with an $\mathcal{ABCD}$-matrix analysis and a numerical simulation. Where the error-bars are not shown they are smaller
than the symbols.}\end{center}
\end{figure}

Two methods of predicting the expected cloud size were employed.  First, an $\mathcal{ABCD}$-matrix analysis was carried out \cite{Gor, focus1, focus2},
characterising the lens as being perfectly parabolic with a finite-duration impulse.  This analysis is easy to perform, but as was pointed out
in~\cite{focus1}, the limit of the validity of the assumptions underlying this method are unlikely to extend to a realistic experiment. The second method
is a brute-force numerical simulation of the trajectories of many atoms subject to the forces of gravity and a pulsed Stern-Gerlach force. In this model,
the magnetic field was calculated by taking the baseball coil to be constructed from eight equal-length, straight, infinitesimally-thin conductors, which
ignored the finite extent of the conductors in the $3\times$3 array of the real lens.  The bias coils were modelled as single current loops, rather than
the 2-turn coils in the experimental lens. Further details of the numerical simulation can be found in~\cite{focus1}. Figure~\ref{2ndpanel} compares the
experimentally obtained cloud size along the $x$-direction with the matrix and numerical simulations.

For the matrix analysis, the initial cloud position and velocity standard deviations are required as input - these were deduced from experimental
measurements.  It is then possible to obtain analytic predictions for the cloud-size dependence on $\tau$ as a function of $\omega_{x}$.  A least-squares
comparison of the data and matrix prediction were made, and the results are summarised in table~\ref{tab:omegax}. The values for $\omega_{x}$ deduced
from  the experimental data are seen to be in good agreement with those predicted from knowledge of the geometry and currents passed through the baseball
lens.

\begin{table}[!b]
\begin{center}
\begin{tabular}{|c|c|c|}
\hline Lens & Predicted $\omega_{x}$ (rad$\,$s$^{-1}$) & Fitted $\omega_{x}$
(rad$\,$s$^{-1}$) \\
\hline \hline
1 & $30\pm1$  & 33 \\
2  & $38\pm1$ & 39  \\
3 & $41\pm2$ & 44 \\
4 & $50\pm2$ & 50 \\
\hline
\end{tabular}
\end{center} \caption{Predicted and fitted values of $\omega_{x}$, the
radial angular frequency.}\label{tab:omegax}
\end{table}

Although the radial frequencies deduced are in good agreement with the expected values, the matrix analysis consistently predicts minimum cloud sizes
which are smaller than those measured experimentally.  The numerical analysis is seen to show far better agreement with the minimum cloud size.  This
confirms the predictions presented in~\cite{focus1} that aberrations arising from terms beyond the ideal parabolic lens approximation are significant.
It should also be noted that the optical pumping of atoms into the state with the largest magnetic moment will not have been complete; consequently some
atoms with smaller magnetic moments will have experienced a smaller magnetic impulse, and will contribute to the larger-than-expected minimum cloud size.

\begin{figure}[!ht]
\begin{center}
\vspace{-5mm} \epsfxsize=.7 \columnwidth\epsfbox{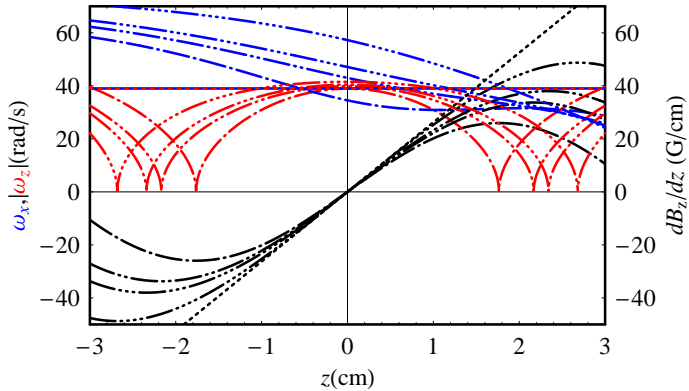} \vspace{-3mm}\caption{\label{curvy} Plots of the axial field gradient (black, right axis),
radial lens frequency and magnitude of the axial lens frequency (blue and red respectively, left axis) as a function of axial position. Configurations 1
to 4 are denoted by dashing with 1 to 4 dots, respectively. The axial lens curvature becomes negative at $z\approx\pm 2$~cm. The gradient (black dots)
and radial and axial angular frequencies (blue and red dots respectively) of an ideal parabolic lens are shown for comparison.}
\end{center}
\end{figure}

A similar analysis was performed for the size of the cloud measured along the $z$-direction. However,  these data do not show good agreement with either
the matrix or numerical simulations, and we now clarify the reasons for this unexpected behaviour.  Figure~\ref{curvy} shows the form of the axial field
gradient, the radial angular frequency and the magnitude of the axial angular frequency as a function of axial position for our lens.  The axial field
gradient is zero at the centre and finite along the axis.  A finite value gives the cloud an additional impulse, which shifts the centre of the image
slightly. It is seen that the radial curvature changes very little over the extent of the cloud, whereas there is considerable variation of the axial
curvature; indeed, the axial curvature changes sign approximately 2~cm from the centre of the lens.  Aligning the centre of the cloud radially with
respect to the centre of the lens is significantly easier than aligning  the centre axially.  As expected, numerical simulations show that the focusing
properties in the $z$-direction are far more sensitive to slight misalignment of the centre of the cloud with respect to the centre of the lens than for
the radial direction. There is also evidence (see below) that the extent of the cloud is longer axially than radially, which will also cause an average
over the cloud of the axial focussing properties of the lens.  We believe that these reasons explain the poor axial performance of the lens.

It is possible to infer from figure~\ref{1stpanel} the properties of the cloud in the limit of the pulse duration, $\tau$, going to zero.  The asymptotic
values radially are in good agreement with what is expected from a ballistically expanding sample of cold atoms at a temperature of 25~$\mu$K, whereas
the axial size is consistently more than 2 times larger.  This suggests the implementation of moving-molasses perhaps caused axial heating of the cloud.

The optimum volume compression of 1/60 is obtained with (e.g.) configuration 1 at $\tau=20\,$ms; here we estimate spatial expansions of 2.9 and 7.3 in
the radial and axial dimensions of the cloud, respectively. Although this is significantly worse than the performance of a harmonic lens where a relative
volume compression of 2.0 is expected,\footnote{This can be obtained from the magnification $(1-\lambda)/\lambda$, where $\lambda$ is the relative time
of the magnetic impulse compared to the atomic time of flight - see \cite{focus1} for further details.} it is two orders of magnitude better than the
unfocused cloud. Liouville's theorem states that the 6 dimensional phase-space density is conserved with pulsed magnetic focusing. However, in
applications such as loading traps and guides a more relevant quantity is the phase-space density of a recaptured rethermalised cloud in a mode-matched
trap. We estimate the experimental phase-space density decrease of the magnetically focused cloud to be two orders of magnitude larger than that which
could be obtained using an ideal parabolic lens. Theoretical simulations with experimentally realistic parameters indicate a decrease by one order of
magnitude relative to the ideal parabolic lens.

\section{Discussion and conclusion}\label{conclusions}

In addition to a study of atom focusing, the results presented here demonstrate a method of transferring  atoms from a MOT to a remote vacuum chamber.
Many cold atom experiments employ a double-chamber system where the first  chamber generally employs a high pressure ($\sim10^{-9}$~Torr) MOT to collect
a large number of cold atoms which are subsequently  transported to a lower pressure `science' chamber to allow for longer trap lifetimes. The act of
moving the atoms between the two regions results in an undesired density decrease unless steps are taken to counteract the cloud's ballistic expansion.
One approach is to catch atoms launched into the science chamber in a second MOT. However, an undesirable feature is the restriction placed on subsequent
experiments by the laser beams and magnetic-field coils required to realise the second MOT. An alternative approach is to focus or guide the launched
atoms such that they can be collected in a conservative trap. Atomic confinement  in the transfer process has been realised both with optical and
magnetic forces. Laser guiding between chambers has been achieved  in free space \cite{Szymaniek99,Pruvost99,Noh02,dimova07}, within optical fibers
\cite{Fibers}, and a BEC has been transported  with an optical tweezer~\cite{Gustavson02}. In a second category of experiments  atoms are loaded into a
magnetic trap in the first chamber, and transported    using either time-dependent currents in an array of static coils \cite{Greiner01}, or trap coils
mounted on a motorised stage \cite{Magtransport}. The disadvantage of a scheme with static coils is the large number of coils and power supplies
required, and the time-dependent currents. Initial experiments with moving coils used a three-dimensional quadrupole trap, which has a magnetic zero at
its centre. For certain applications, a trap with a finite minimum is required, and recently transport of atom packets in a train of Ioffe-Pritchard
traps was demonstrated \cite{Lahaye06}. Using moving coils does, however, place limitations on vacuum chamber design since sufficient space must be
allowed for the translation mechanism. The advantage of the pulsed magnetic lens presented in this work is that only space for the lens itself is needed.
However, in contrast to this work, there is no significant increase in cloud size with the two magnetic transport schemes described above. The
combination of pulsed magnetic focusing combined with laser guiding also looks promising \cite{transport06}.

In this work spatial focusing was considered.  A possible future extension would be to study velocity focusing, and recently a Ioffe-Pritchard lens was
used for this purpose~\cite{aoki06}.  A wavepacket with a very narrow momentum distribution is ideal for studying quantum tunnelling, and a 1-dimensional
narrow momentum distribution could also be useful for other atom optics experiments, such as studying quantum accelerator modes.

\ack This work is supported by EPSRC. We thank Charles Adams and Simon Cornish for fruitful discussions.  Kevin Weatherill designed the external-cavity
laser.

\end{document}